\documentstyle[aps,multicol,epsfig]{revtex}
\draft \tolerance=10000

\def\vec#1{{\rm\bf #1}}

\def\lz{\ell_{\parallel}}
\def\lp{\ell_{\bot}}
\def \be{\begin{equation}}
\def \bea{\begin{eqnarray}}
\def \ee{\end{equation}}
\def \eea{\end{eqnarray}}
\newcommand{\tens}[1]{\underline{\underline{#1}}}

\begin{document}
\title{Untwisting of a cholesteric elastomer by a mechanical field}
\author{M.~Warner, E.~M.~Terentjev, R.~B.~Meyer${}^*$,
Y.~Mao}
\address{Cavendish Laboratory, University of Cambridge,
Madingley Road, Cambridge CB3 0HE, U.K.\\ ${}^*$ The Martin Fisher
School of Physics, Brandeis University, Waltham, MA 02454-9110,
USA}
\date{\today}
\maketitle
%%%%%%%%%%%%%%%%%%%%%%%%%%%%%%%%%%%%%%%%%%%%%%%%%%%%%%%%%%%%%%%
\begin{abstract}
A mechanical strain field applied to a monodomain cholesteric
elastomer will unwind the helical director distribution. There is
an analogy with the classical problem of an electric field applied
to a cholesteric liquid crystal, but with important differences.
Frank elasticity is of minor importance unless the gel is very
weak. The interplay is between director anchoring to the rubber
elastic matrix and the external mechanical field. Stretching
perpendicular to the helix axis induces the uniform unwound state
via the elimination of sharp, pinned twist walls above a critical
strain. Unwinding through conical director states occurs when the
elastomer is stretched along the helical axis.
\end{abstract}
%%%%%%%%%%%%%%%%%%%%%%%%%%%%%%%%%%%%%%%%%%%%%%%%%%%%%%%%%%%%%%%
 \vspace{0.2cm}
%\narrowtext
\begin{multicols}{2}
%%%%%%%%%%%%%%%%%%%%%%%%%%%%%%%%%%%%%%%%%%%%%%%%%%%%%%%%%%

Monodomain cholesteric elastomers are formed by crosslinking
mesogenic chiral polymers in the cholesteric state with a properly
formed helical director twist. The subsequent retention of the
helical state as an elastic equilibrium \cite{zentel} is a natural
consequence of topological imprinting of textures in the
crosslinked network, seen in a number of other elastomers with
liquid crystalline order and other microstructure. Recently an
interesting aspect of chiral imprinting was established by
crosslinking nematic polymers in a chiral state purely induced by
a chiral solvent \cite{geof}; on removal of the solvent, the
network of chemically achiral nematic chains remains
macroscopically cholesteric. Such an imprinting has been envisaged
long time ago \cite{degennes} on phenomenological grounds. It is
now important to consider the mechanical possibilities of such
solids with a helical microstructure, expecting new transitions
and instabilities characteristic of liquid crystalline elastomers.
Additionally there are obvious device applications of such
materials, which combine all the optical properties of twisted
nematic liquids with the remarkable mechanical characteristics of
rubbers. There is some experimental evidence \cite{zent2} that
such effects are indeed observable and our hope is that this
theoretical work will stimulate more studies in this field.

%%%%%%%%%%%%%%%%%%%%%%%%%%%%%%%%%
\begin{figure}[h]
\centerline{ \epsfxsize=7cm \epsfbox{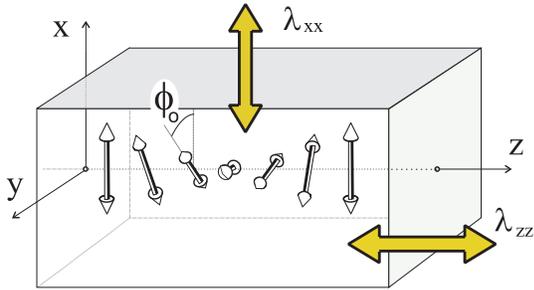}} \vspace{0.25cm}
 \caption{The
initial director $\vec{n}_0(z)$ in a cholesteric helix makes an
azimuthal angle $\phi_0 = q_0z$ with the $x$-axis; the helical
pitch is $p = \pi/q_0$. Two principal directions of mechanical
deformation, $\lambda_{\rm xx}$ and $\lambda_{\rm zz}$, are shown
by arrows.} \label{coordinates}
\end{figure}
%%%%%%%%%%%%%%%%%%%%%%%%%%%%%%%
Consider a monodomain cholesteric elastomer with an ideal
helically twisted director $\vec{n}_0(z)$ in the $xy$ plane,
initially making angle $\phi_0=q_0z$ with the $x$ axis,
Fig.~\ref{coordinates}. We shall examine two specific cases of
imposed uniaxial extension: (i) the transverse deformation
$\lambda_{\rm xx}=\lambda$, in the plane including $\vec{n}_0$,
and (ii) the longitudinal deformation along the helix axis
$\lambda_{\rm zz}=\lambda$.

The symmetry obvious from Fig.~\ref{coordinates} requires that in
the case (i) the director remains in the $xy$ plane, characterised
by the azimuthal angle $\phi(z)$, while in the case (ii) one may
expect a conical texture with $\vec{n}(z)$ inclined towards the
stretching axis $z$ and, therefore, described by two angles
$\theta$ and $\phi$ (cf. Fig.~\ref{cone} below). In ordinary
liquid cholesterics subjected to, e.g., a magnetic field $H_z$,
such conical states are not generally seen, preempted by the
$90^{\rm o}$-switching of the helix axis and then untwisting in
the ``transverse'' geometry \cite{meyer}. We shall see that in
elastomers, due to the chiral imprinting, this regime is not
possible and the conical director configurations should occur.

An elastic material with a microstructure represented by an
independently mobile director orientation is analogous to the
Cosserat medium. In the limit of linear elasticity the relative
rotation coupling between the director rotation
$\bbox{\omega}=[\vec{n}\times \delta \vec{n}]$ and the
antisymmetric part of strain, $\Omega_{\rm i} = \epsilon_{\rm ijk}
\varepsilon_{\rm jk}$,
\be
\frac{1}{2}D_1 [\vec{n} \times (\bbox{\Omega - \omega})]^2 + D_2
\vec{n}\cdot \tens{\varepsilon}^{(s)} \cdot [\vec{n} \times
(\bbox{\Omega - \omega})] ,  \label{D12}
 \ee
has been first written down phenomenologically by de Gennes
\cite{d12}, $ \tens{\varepsilon}^{(s)}$ being the symmetric part
of the small strain defined as
$\tens{\varepsilon}=\tens{\lambda}-\tens{\delta}$. This
symmetry-based expression is only valid for small deformations,
having only linear and quadratic terms in the local relative
rotation.

 The microscopic statistical-mechanical theory of nematic rubber
elasticity, e.g. \cite{wt}, obtains a generalisation of the
classical rubber-elastic energy density in the form of a complete
frame-independent expression
\be
F=\frac{1}{2} \mu \ {\sf Tr}\left(\tens{\ell}_0 \cdot
\tens{\lambda}^T \cdot \tens{\ell}^{-1} \cdot
\tens{\lambda}\right), \label{F0}
 \ee
plus the constraint of material incompressibility, expressed by
the condition ${\sf Det} \left( \tens{\lambda} \right)=1$ on the
strain tensor. Apart from the strain tensor, the other entries in
the Eq.~(\ref{F0}) are $\tens{\ell}_0=\lp \tens{\delta}
+(\lz-\lp)\vec{n}_0\,\vec{n}_0$ and $\tens{\ell}^{-1}=(1/\lp)
\tens{\delta} +(1/\lz - 1/\lp )\vec{n}\,\vec{n}$, the reduced
shape and inverse shape tensors characterising the Gaussian
distribution of nematic polymer chains before and after the
distortion $\tens{\lambda}$. The rubber shear modulus $\mu=n_s
k_BT$ (with $n_s$ the number density of network strands,
proportional to the crosslink density) is that characteristic of
the underlying isotropic rubber and sets the energy scale of
distortions. The free energy density Eq.~(\ref{F0}) is known to be
valid up to large strains and correctly predicts the
opto-mechanical responses and the soft elasticity of nematic
elastomers. The free energy $F$ is a function only of the chain
anisotropy $r=\lz/\lp$, the ratio of the effective step lengths
parallel and perpendicular to the director. It is an independently
measured parameter accessible from neutron scattering or from
spontaneous mechanical distortions on going from the nematic to
isotropic phase. Unless there is a large nematic order change
induced by $\tens{\lambda}$, the shape $\tens{\ell}$ is
essentially just a rotated version of $\tens{\ell}_0$, a uniaxial
ellipsoid with the long axis (at $r>1$) oriented along $\vec{n}$
instead of $\vec{n}_0$.

Embedded in the general expression Eq.~(\ref{F0}) is the penalty
for local director deviations from the orientation $\vec{n}_0$
imprinted into the network at formation. When no elastic strains
are allowed, this elastic energy reduces to
\be
F \rightarrow \frac{3}{2}\mu + \frac{1}{2} \mu \frac{(r-1)^2}{r}
\sin^2 \Theta \label{D1}  \ee
 where $\Theta$ is the local angle between $\vec{n}$ and
$\vec{n}_0$. The elastic penalty for such a deviation,
appropriately proportional to the square of chain anisotropy, is
the coefficient $D_1$ of the de Gennes' phenomenological
expression at small deformations, Eq.(\ref{D12}). This has to be
compared with the Frank elastic penalty for director curvature
deformations, $\frac{1}{2}K (\nabla \vec{n})^2$. The length scale
$\xi \sim \sqrt{K/\mu}$ at which the two energy contributions are
comparable is usually small: $\xi \sim 10^{-8}$m for a typical $K
\sim 10^{-11}\hbox{J/m}, \, \mu \sim 10^5\hbox{J/m}^3$ and not too
small anisotropy, $r$. This is rather less than the cholesteric
pitch $p$, which is a characteristic scale in our problem.
Therefore, the anchoring of the director $\vec{n}$ to the rubbery
matrix, described by Eq.~(\ref{F0}), tends to dominate over Frank
effects.

We shall assume that a cholesteric elastomer is locally like a
nematic in its elastic response: rubber elasticity is determined
on the scale of network crosslink separations (a few nanometers),
whereas cholesteric pitches are $10^3$ times longer. We can at
once see why the chiral structure is stable and how mechanical
fields can destabilise it. With no elastic strain, the free energy
penalty is $\sim \textstyle{1\over 2} D_1 (\phi-\phi_0)^2$ for
rotating the director away from its original helical texture
$\phi_0=q_0z$. On the other hand, if strains are applied, the
rubber can lower its elastic energy Eq.~(\ref{F0}) by rotating the
director $\vec{n}$ towards the axis of principal extension. This
general principle of adjusting the microstructure to minimise the
elastic energy is seen in its ultimate form in the effect of soft
elasticity \cite{wt,olm}, when a stretched nematic rubber may
reduce its effective modulus (the slope of a stress-strain curve)
to zero by optimising the director rotation and associated shear
strains.

Distortions in a cholesteric elastomer cannot be soft, because of
elastic compatibility constraints in matching different director
and shear modes along the helix. If the director at a position $z$
rotates towards the $x$-axis, it is known that the elongation
$\lambda_{\rm xx}$, contraction $\lambda_{\rm yy}$, and shear
$\lambda_{\rm xy}$, are precisely determined by the initial
orientation, $\phi_0$, and the rotation from it, if the process is
to be soft. The next slab of material, at $z+dz$, has the initial
orientation $\phi_0+q_0dz$ and a different set of soft strains
$\tens{\lambda}$ must arise. Material points at $y$ translate to
$\lambda_{\rm yy}(z)\cdot y$ and $\lambda_{\rm yy}(z+dz) \cdot y$
in the two neighbouring slabs along the helix, that is they differ
by a relative displacement $(\partial \lambda_{\rm yy} /
\partial z) dz \cdot y$. There is thus a generated shear
$\lambda_{\rm yz}=(\partial \lambda_{\rm yy}/\partial z)\cdot y$
that diverges as the linear $y$-dimension of the sample. We
accordingly conclude that the transverse contractions are {\it
uniform}. Such deformations, e.g. $\lambda_{\rm zz}$ and
$\lambda_{\rm yy}$, are equal in the first approximation, in spite
of the apparent anisotropy along the pitch axis $z$ (each
deformation generates the same energy $k_BT$ per chain
independently of its alignment with respect to the director) and
are thus both equal to $1/\sqrt{\lambda}$.

\begin{figure} %[t]
\centerline{ \epsfxsize=0.25\textwidth \epsfbox{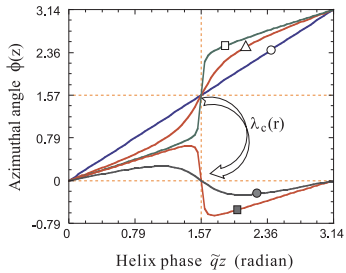}}
\vspace{0.25cm}
 \caption{The director angle $\phi$ against the
cholesteric helix phase $q_0z$ for increasing strain $\lambda$ = 1
(open circle), 1.15 (triangle), 1.23 (open square), 1.25 (shaded
square) and 1.5 (shaded circle). The polymer chain anisotropy is
$r=1.9$ and thus the critical strain is $\lambda_c = r^{1/3}
\approx 1.24$ (see text). At $\lambda > \lambda_c$ the director
pinning at $\phi=\pi/2$ breaks down and a discontinuous transition
occurs, after which the director continuously rotates towards the
final uniform $\phi=0$.} \label{transverse}
\end{figure}

(i) {\sc Transverse elongation}  $\lambda_{\rm xx}=\lambda$. We
consider the strain tensor in the following form:
\be
\tens{\lambda}=\left( \begin{array}{ccc} \lambda & 0 & 0\\ 0 &
1/\sqrt{\lambda} & 0\\ 0 & 0 & 1/\sqrt{\lambda}\\
\end{array} \right).
 \ee
Although one expects the director rotation in the azimuthal plane
(cf. Fig.~\ref{coordinates}), there are no associated shear
strains. Such shears, $\lambda_{\rm xy}(z)$ and $\lambda_{\rm
yx}(z)$, would both lead to elastic compatibility problems (e.g. a
generated shear $\lambda_{\rm xz} \propto y$, the sample
dimension) and we assume they are suppressed. The shears
$\lambda_{\rm xz}(z)$ and $\lambda_{\rm xz}(z)$ are not subject to
compatibility requirements. However, they should not appear on
symmetry grounds, which is easily confirmed by direct
minimisation. Now $\vec{n}_0=\{ \cos \, \phi_0, \, \sin \, \phi_0
, \, 0\}$ and the rotated director after deformation is $\vec{n}
=\{ \cos \, \phi, \, \sin \, \phi, \, 0\}$. Note that the helix is
$\phi_0=q_0z$ in the initial undistorted material. After
deformation, because of the affine contraction $\lambda_{\rm
zz}=1/\sqrt{\lambda}$, the material frame shrinks: ${z}\rightarrow
z/\sqrt{\lambda}$. This has an effect of affine contraction of the
helical pitch so that $\tilde{q}=\sqrt{\lambda} \, q_0 $ in all
expressions below. With the $\tens{\ell}_0$ and $\tens{\ell}$
implied by these $\vec{n}_0$ and $\vec{n}$, the free energy
density Eq.~(\ref{F0}) yields
 \bea
F_\bot &=& \frac{1}{2}\mu \left( \lambda^2 + {2 \over \lambda}+
{r-1 \over r}\left[ \lambda^2(r\;c_0^2\;s^2-c^2\;s_0^2)  \right.
\right. \label{Ftransv} \\
 \qquad &&+ \left. \left. {1\over \lambda}(r\; c^2
s_0^2 - s^2 c_0^2) - 2\sqrt{\lambda} (r-1) s_0\;c_0\;s\;c \right]
\right), \nonumber
 \eea
where $c_0$ and $s_0$ are shorthand for $\cos \, \phi_0$ and $\sin
\, \phi_0$; analogously, $c$ and $s$ stand for $ \cos \, \phi$ and
$\sin \, \phi$. The appearance of terms linear and quadratic in
$\phi$ (or rather $\sin \phi$ because all values of the azimuthal
angle will be found along the cholesteric helix) indicate that
rotations can always lower the energy for $\lambda \neq 1$.
Minimization of $F_\bot$ with respect to $\phi$ results in the
expression for the local director angle $\phi(z)$ at a given
imposed extension $\lambda$, depending on the phase of cholesteric
helix:
\be
\tan 2\phi =\frac{2\lambda^{3/2} (r-1) \sin \, 2 \tilde{q}z}{
(r-1)(\lambda^3+1)\cos \,2 \tilde{q}z +(r+1)(\lambda^3-1) }
\label{transv}
 \ee
Initially, all directors at $0< \tilde{q}z<\pi/2$ are induced to
rotate ``backward'' towards $\phi=0$, and all directors at
$-\pi/2<  \tilde{q}z < 0$ rotate ``forward'' towards $\phi= \pi$,
as the imposed deformation $\lambda$ increases, see
Fig.~\ref{transverse}. Although $\phi=0$ and $\pi$ describe
equivalent directors, the twist wall between these two states
becomes more and more sharp. Due to the helix imprinting, the
orientations $\phi=0$ at $ \tilde{q}z=0$ and $\phi=\pi$ at $
\tilde{q}z=\pi$ are pinned, as is the middle-point of the twist
wall at $ \tilde{q}z=\pi/2$. As a result, no change of the helical
pitch can occur. This is in contrast with cholesteric liquid
crystals, where in a classical problem of helix unwinding by
electric or magnetic field one finds an increase in cholesteric
pitch \cite{meyer,dgwinding} along with the coarsening of the
helix.
%%%%%%%%%%%%%%%%%%%%%%%%%%%%%%%%%
%%%%%%%%%%%%%%%%%%%%%%%%%%%%%%%

Examining the Eq.(\ref{transv}) one finds that the denominator
changes sign and remains negative in the region of the twist wall,
centered at $ \tilde{q}z=\pi/2$ between the values $\phi=\pi/4$
and $3\pi/4$. The width of such a wall is $$w\simeq \frac{2}{q_0}
\sqrt{\frac{r-\lambda^3}{(r-1) (\lambda^3+1)}}.$$ As the
increasing applied strain reaches a critical value $\lambda_c=
r^{1/3}$, the wall width $w\rightarrow 0$ and the discontinuous
transition occurs. The director in the mid-point of the wall
breaks away from the pinning and jumps from $\phi=\pi/2$ to
$\phi=0$, along the strain axis, thus removing the topologically
constrained twist wall. From this point there is no barrier for
director rotation towards the final uniform orientation with
$\phi=0$, as the last two curves in Fig.~\ref{transverse}
indicate. The bifurcation occurs between the family of $\phi(z)$
curves at $\lambda <\lambda_c$ and those at $\lambda>\lambda_c$.
They are separated by the critical curve $\phi_c$ given by $\tan
2\phi_c = - 2\left( \frac{\sqrt{r}}{r+1} \right) \tan \,
\tilde{q}z$.

A discontinuous director jump at a critical strain has been
predicted, and indeed observed in nematic elastomers stretched at
$90^{\rm o}$ to their initial director $\vec{n}_0$ \cite{wt,GRM}.
It was also found that even small deviations from the exact
perpendicular geometry, or the possibility of soft shears in
stripe domains lead to a continuous director rotation. However, in
the present cholesteric case the shears are prohibited by
compatibility constraints and one always finds an exact phase
angle $\phi=\pi/2$ along the helix -- where the center of
narrowing twist wall becomes pinned from both sides. It is this
point that experiences a discontinuous jump in order to allow the
system to proceed to its final uniform equilibrium state.

(ii) {\sc Stretching along the pitch axis} $\lambda_{\rm
zz}=\lambda$. We now take (cf. Fig.~\ref{coordinates})
\be
\tens{\lambda}=\left( \begin{array}{ccc} 1/\sqrt{\lambda} & 0 &
\lambda_{\rm xz}\\ 0 & 1/\sqrt{\lambda} & \lambda_{\rm yz}\\ 0 & 0
& \lambda \end{array} \right)
 \ee
No compatibility problem with shears $\lambda_{\rm xz}(z)$ and
$\lambda_{\rm yz}(z)$ arises from their variation with $z$ along
the helical pitch. By contrast, their conjugate strains
$\lambda_{\rm zx}$ and $\lambda_{\rm zy}$, which would also have
to vary with $z$, would lead to a serious compatibility mismatch,
e.g. $\partial \lambda_{\rm zx}/\partial z =\partial \lambda_{\rm
zz}/\partial x$. We therefore assume $\lambda_{\rm zx}$ and
$\lambda_{\rm zy}$ are suppressed even though in other settings
\cite{wt} these are the generators of soft elastic response.
%%%%%%%%%%%%%%%%%%%%%%%%%%%%%%%%%
\begin{figure} %[b]
\centerline{ \epsfxsize=0.45\textwidth \epsfbox{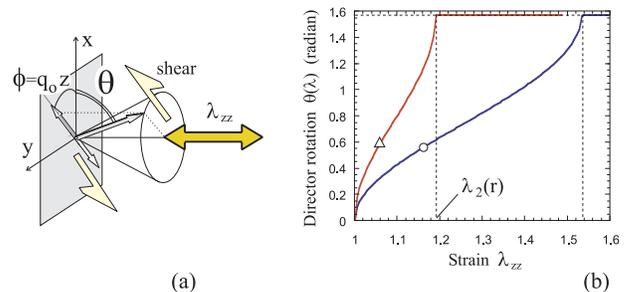}}
\caption{ (a) The geometry of director rotation in response to
stretching $\lambda_{\rm zz}$ along the helix axis. \ (b)  The
angle $\theta$ of director tilt plotted against the imposed strain
$\lambda$, Eq.~(\ref{tilt}) for $r=1.3$ (triangle) and $r=1.9$
(circle). Strain varies from 1 to $\lambda_2=r^{2/3}$ at which
point the alignment is $\theta=\pi/2$, uniformly along the former
pitch axis.} \label{cone}
\end{figure}
%%%%%%%%%%%%%%%%%%%%%%%%%%%%%%%

In this geometry one expects the director rotation out of the
azimuthal $xy$ plane, see Fig.~\ref{cone}(a). The initial director
is, as before, $\vec{n}_0=\{ \cos \, q_0z, \, \sin \, q_0z , \,
0\}$, while after deformation the  rotated director is $\vec{n}=\{
\cos \theta \, \cos \,  \tilde{q}z, \, \cos \theta \, \sin \,
\tilde{q}z, \, \sin \theta \}$. As in the case (i), all physical
dimensions in the deformed sample are scaled by the affine strain.
In particular, here ${z} \rightarrow \lambda z$, resulting in the
corresponding expansion of the cholesteric pitch: $\tilde{q} =
q_0/\lambda$. With the $\tens{\ell}_0$ and $\tens{\ell}$ defined
by the axes $\vec{n}_0$ and $\vec{n}$, the free energy density
Eq.~(\ref{F0}) now becomes a function of three variables: the
director tilt angle $\theta$ and the two shear strains
$\lambda_{\rm xz}(z)$ and $\lambda_{\rm yz}(z)$ (we continue to
neglect the effects of director gradients and Frank elasticity).
Algebraic minimisation over these components of strain tensor is
not difficult and results in
\be
\left( \begin{array}{c} \lambda_{\rm xz} \cr \lambda_{\rm yz}
\end{array} \right) = \lambda \frac{(r-1)\sin \,
2 \theta}{r+1-(r-1)\cos \, 2\theta}\left( \begin{array}{c} \cos \,
\tilde{q}z \cr \sin \, \tilde{q}z \end{array} \right) ,
\label{shears}
 \ee
in phase with the azimuthal angle along the helical pitch.
Eq.~(\ref{shears}) describes small distortions in the $xy$ plane,
perpendicular to the helix axis, rotating following the initial
orientation $\vec{n}_0$. On substitution of these optimal shears
back into the free energy density one obtains
\be
F_\|=\frac{1}{2}\mu \left( \frac{\lambda^2}{1+(r-1)\sin^2 \theta}
+ \frac{2+(r-1)\sin^2 \theta}{\lambda}  \right) \label{Fpar}
 \ee
$F_\|$ expands at small tilt angle $\theta$ as
\be
F_\| \approx {1 \over 2}\mu (\lambda^2+2/\lambda)- {1 \over 2}\mu
\, \theta^2(r-1)(\lambda^2-1/\lambda), \nonumber
 \ee
that is the director starts to rotate down to define a cone of
semiangle $\pi/2 -\theta$ immediately as the strain $\lambda>1$ is
imposed. The equilibrium director tilt is obtained by minimisation
of the full free energy density $F_\|(\theta)$:
\be
\sin^2 \theta ={\lambda^{3/2}-1 \over r-1}; \qquad \theta=\arcsin
\sqrt{\lambda^{3/2}-1 \over r-1}\; . \label{tilt}
 \ee
The director rotation starts and ends in a characteristically
singular fashion Fig.~\ref{cone}(b) (reminiscent of the universal
opto-mechanical response seen in nematic elastomers
\cite{finkelmann}). The rotation is complete with the director
aligned along the extension axis ($\theta=\pi/2$) at
$\lambda=r^{2/3}$ which, for some elastomers, can be a very large
extension.

In contrast to conventional cholesteric liquid crystals, we have
altogether ignored effects of Frank elastic energy. The most
compelling evidence for this is the very stability of the
imprinted helical state in the face of the Frank penalty
$\textstyle{1\over 2} K_2 q_0^2$. The argument for this relies
upon the great difference in characteristic length scales, the
elastomer penetration depth, more accurately expressed as $\xi
\simeq \frac{1}{r-1}\sqrt{K/\mu}$ [cf. Eq.~(\ref{D1})], and the
director modulation wavelength estimated by the helical pitch
$p=\pi/q_0 \gg \xi$. There are two possibilities to alter this
inequality -- by increasing the penetration depth $\xi$ (either by
making a weaker gel, or a less anisotropic one), or by locally
increasing the director gradient (for instance, in the ever
narrowing twist wall, Fig.~\ref{transverse}).

One can estimate how weak a gel must be for the Frank elasticity
to intervene in our analysis. When $\xi \sim p$, for example with
a pitch $p \sim 4 \times 10^{-7}$m, then a rubber modulus of only
$\mu \sim 60 \, \hbox{J/m}^3$ is required (assuming $[r-1]\sim
1$). Nematic elastomers typically have $\mu \sim 10^3 - 10^5
\hbox{J/m}^3$ and their cholesteric analogues would clearly find
Frank-elastic effects minor. However, an elastomer with a
reasonable $\mu \sim 10^3 \hbox{J/m}^3$ would feel the director
gradients when its polymer chain anisotropy becomes as low as
$r=\lz/\lp \sim 1.25$. Such a value is easily reached in
side-chain liquid crystal polymers, especially near the clearing
point \cite{wt}.

Another interesting test of the role of Frank elasticity is in the
twist wall described in our case (i). The width of the wall
decreases to zero, and the Frank energy density grows, being
maximal at the centre of the wall. There it is
$\frac{1}{2}K_2q_0^2 \lambda^3(r-1)^2/(r-\lambda^3)$ and diverges
at the critical strain $\lambda_c$. Therefore, the local analysis
of Eqs.~(\ref{Ftransv})-(\ref{transv}) is only valid outside the
region of strain $\Delta \lambda \approx \left(
\frac{r^2-1}{3r^{2/3}} \right) (q_0\xi)^2$ around $\lambda_c$. In
a typical hard nematic rubber this is a narrow region of $\Delta
\lambda \leq 0.01$, but in a weak gel with low chain anisotropy it
may become more substantial. Moreover, the finite width of the
twist wall, demanded by the Frank gradient energy, raises the
question of topological mechanism for eliminating the twist stored
in the cholesteric helix, perhaps by a disclination loop
expansion.

To summarise, we have predicted a qualitatively new response of an
elastomer with chiral cholesteric microstructure to applied fields
that is different from classical cholesteric liquids. Likewise,
the chiral imprinting and its modification by elastic fields is a
new effect in rubbers and solids. One could envisage tuning these
effects by the use of solvents (with or without chiral power) and
by other fields affecting the director, for instance electric. \\

%\section*{Acknowledgement}
YM is grateful to St John's College, Cambridge for a research
fellowship and RBM acknowledges support by the NSF, through grant
DMR-9974388.

%%%%%%%%%%%%%%%%%%%%%%%%%%%%%%%%%%%%%%%%%%%%%%%%%%%%%%%%%%%%%%%%%%%
%\begin{thebibliography}{99}

%\end{thebibliography}
%%%%%%%%%%%%%%%%%%%%%%%%%%%%%%%%%%%%%%%%%%%%%%%%%%%%%%%%%%%%%%%%%%%
\end{multicols}
\end{document}